\documentclass[preprint]{aastex}

\begin{document}


\slugcomment{accepted by ApJL: December 9, 2010}

\title{
Millimeter Imaging of the $\beta$~Pictoris Debris Disk: \\
Evidence for a Planetesimal Belt 
}

\author{
David J. Wilner\altaffilmark{1},
Sean M. Andrews\altaffilmark{1},
A. Meredith Hughes\altaffilmark{2,3}
}
\altaffiltext{1}{Harvard-Smithsonian Center for Astrophysics, 60 Garden Street, Cambridge, MA 02138}
\altaffiltext{2}{Department of Astronomy, 601 Campbell Hall, University of 
California, Berkeley, CA 94720} 
\altaffiltext{3}{Miller Fellow}

\begin{abstract}
We present observations at 1.3~millimeters wavelength of the 
$\beta$~Pictoris debris disk with beam size
$4\farcs3 \times 2\farcs6$ ($83\times50$~AU)
from the Submillimeter Array. 
The emission shows two peaks separated by $\sim7''$ along the disk 
plane, which we interpret as a highly inclined dust ring or belt. 
A simple model constrains the belt center to $94\pm8$~AU, close to the 
prominent break in slope of the optical scattered light. We identify 
this region as the location as the main reservoir of dust producing 
planetesimals in the disk.

\end{abstract}


\section{Introduction}
The discovery of excess far-infrared emission from the 
nearby \citep[$19.44\pm0.05$~pc;][]{van07} 
A6V-type main-sequence star $\beta$~Pictoris \citep{aum85} together with 
optical imaging of scattered light from circumstellar dust \citep{smi84} 
established the ``debris disk'' paradigm where dust grains orbiting 
the star originate from an eroding reservoir of larger bodies
\citep[see reviews by e.g.][]{art97,bac93,wya08}.
The nearly edge-on disk surrounding this young star
\citep[$12^{+8}_{-4}$~Myr;][]{zuc01} is relatively luminous
\citep[$F_{disk}/F_{*} = 2.5 \times 10^{-3}$;][]{lag00} and has been 
studied in great detail with a panoply of observational techniques. 
High resolution images in the optical \citep{kal95,hea00,gol06},
near-infrared \citep{mou97b,tam06,boc09}
and mid-infrared \citep{wah03,wei03,oka04,tel05}
all show a wealth of structure, including density 
concentrations, an inner cavity, and asymmetries such as warps.
These features, including a secondary disk of scattered light 
inclined by about $5{\degr}$ \citep{ahm09}, have been variously ascribed 
to the gravitational influence 
of a giant planet or planets \citep[e.g.][]{mou97a,aug01,fre07,ken10}.
Indeed, a planetary mass companion at a projected distance of 8~AU 
from the star now has been directly imaged \citep{lag09,lag10}.

The emerging view of debris disks like $\beta$~Pictoris postulates 
the presence of a planetesimal belt that produces dust with a range of 
sizes through collisional cascades \citep{str06,wya08,kuc10}. 
The stirring of the planetesimals may be due to the gravity of 
$\sim1000$~km-sized objects formed within the belt \citep{ken04}, or to 
the presence of planets located closer to the star \citep{mor07}. 
In either case, the dynamical effects of stellar radiation create 
a distribution of grain sizes that depends on distance from the star, 
e.g. the blow-out of the smallest ``$\beta$-meteoroid'' grains. 
An important consequence is that images of debris disks at different 
wavelengths are dominated by different grain sizes and can appear 
remarkably different \citep{wya06}. 
Observations at millimeter wavelengths are most sensitive to large 
grains that are minimally affected by radiative forces and thus 
have the potential to trace best the location of 
the dust producing parent planetesimals. 
The debris disk around Vega, for example, shows a clumpy ring confined 
to radii $<200$~AU at wavelengths of 350~$\mu$m  and longward
\citep{hol98,wil02,mar06}, while it appears smooth and featureless and 
extends to radii $\sim800$~AU in mid-infrared light that arises predominantly 
from small grains expelled by radiation \citep{su05}. 
The debris disk around HR~8799, an A-type star that harbors three 
directly imaged planets, shows similar morphological changes with 
wavelength \citep{su09}.

For the $\beta$~Pictoris debris disk, the angular resolutions of 
(sub-)millimeter images have been too coarse to reveal much structure. 
Images from several different telescopes generally show dust emission extended
along a position angle of $\sim30{\degr}$, consistent with the optical disk:
JCMT/SCUBA at 850~$\mu$m with a $14''$ beam \citep{hol98},
APEX/LABOCA at 870~$\mu$m with an $18''$ beam \citep{nil09}, 
SEST/SIMBA at 1200~$\mu$m with a $24''$ beam \citep{lis03},
and Herschel/SPIRE at 250, 350 and 500~$\mu$m with $18$, $25$, and $37''$ 
beams, respectively \citep{van10}.
A separate peak or blob of dust emission is also found $\sim30''$ to the 
southwest of the star, but the relationship of this peak to the main disk 
is unclear; \citet{den00} and \citet{van10} have suggested it is a 
background galaxy with a coincidental alignment with the disk plane.

Millimeter interferometry offers a way to obtain higher angular resolution
and more information on the largest detectable grain populations within
the debris disk. In this {\em Letter}, we present imaging observations of 
$\beta$~Pictoris at 1.3~millimeters wavelength from the 
Submillimeter Array (SMA)\footnote{The Submillimeter Array is a joint project 
between the Smithsonian Astrophysical Observatory and the Academica Sinica 
Institute of Astronomy and Astrophysics and is funded by the Smithsonian 
Institution and the Academica Sinica.}
that reveal a belt of emission around the star centered near a radius of 
$\sim95$~AU that likely marks a reservoir of planetesimals. 

\section{Observations}
\label{sec:obs}
We used the eight element SMA \citep{ho04} on Mauna Kea, Hawaii to observe 
$\beta$~Pictoris in the compact-north 
configuration (baselines 6-97~m) on 2010 August 9, and in the 
extended configuration (baselines 12-178~m) on 2010 September 1. 
The phase center was chosen to be
$\alpha = 5^h47^m17\fs09$, $\delta =-51\degr03\arcmin59\farcs5$ (J2000),
about $0\farcs9$ from the stellar position at the epoch of the observations.
The $\beta$~Pictoris system is a challenging target for the SMA as it
never rises above $20{\degr}$ elevation. Nonetheless, usable data 
were obtained in both configurations over the hour angle range $\pm1.7$. 
The weather conditions were good on both days, with 225~GHz atmospheric 
opacities 0.07--0.09 and stable atmospheric phase.  
The correlator was configured to provide the maximum 4~GHz of bandwidth 
in each of two sidebands centered $\pm6$~GHz from a central LO frequency 
of 235.6~GHz (wavelength of 1.3~millimeters), with a uniform spectral 
resolution of 0.8125~MHz. 
At this frequency, the primary beam size is $\sim54''$ (FWHM).
Observations of the strong source 3C454.3 obtained at the start of each 
track were used to calibrate the passband response. 
Observations of the quasars J0538-440 and J0522-364 were interleaved 
with $\beta$~Pictoris in order to calibrate time dependent gain variations.
The astrometric uncertainty is $\lesssim 0\farcs3$. 
The absolute flux scale 
was set with reference to observations of the standard calibrator Callisto 
in each track and should be accurate to better than 15\%.
The calibration procedure was performed using the IDL based MIR software.
Subsequent imaging and deconvolution were done within the MIRIAD package.

\section{Results and Analysis}
\label{sec:result}
\subsection{1.3 Millimeter Emission}
\label{sec:continuum}
\begin{figure}[h!]
\begin{center}
\includegraphics[scale=1.00,angle=0]{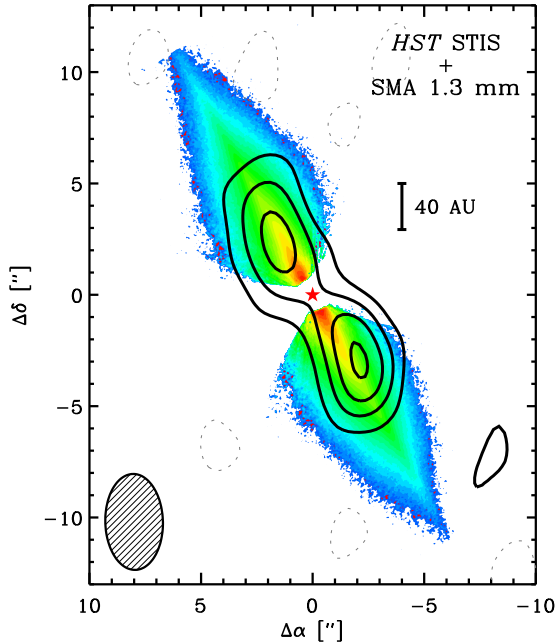}
\figcaption{
SMA image of the 1.3 millimeter continuum emission from $\beta$~Pictoris
overlayed on an image of optical scattered light from \cite{hea00}. 
The contour levels are $-2,2,4,6,...\times0.6$~mJy (the rms noise level),
Negative contours are dotted.
The ellipse in the lower left corner represents the 
$4\farcs3 \times 2\farcs6$ (FWHM) synthesized beam size.
The star symbol indicates the location of the stellar photosphere.
\label{fig:image}}
\end{center}
\end{figure}

Figure~\ref{fig:image} shows a contour image of the 1.3~millimeter emission 
overlaying a Hubble Space Telescope/STIS coronographic image of 
optical scattered light from \citet{hea00}. The 1.3~millimeter 
image was made using natural weighting and a modest taper in the east-west 
direction to avoid extreme ellipticity of the synthesized beam, which is
$4\farcs3 \times 2\farcs6$ ($83\times50$~AU) oriented nearly north-south 
(position angle $2{\degr}$). 
The maximum sidelobes of the dirty beam obtained with this weighting scheme 
are located about $13''$ to the east and west, with amplitude 15\% of the 
central peak.
The rms noise in this image is 0.6~mJy~beam$^{-1}$. 
The star symbol is plotted offset by $(0\farcs35, 0\farcs70)$ from the phase 
center, within the uncertainties of the stellar position corrected for proper 
motion. The 1.3 millimeter emission shows two peaks at positions symmetrically 
offset from the stellar position by $\sim 3\farcs5$ to the northeast and 
southwest, 
respectively. This basic morphology suggests a highly inclined ring or belt, 
where the peaks are due to limb brightening at the ansae (where the column 
density is highest). While the southwest peak appears slightly brighter, the 
difference lies within the noise and cannot be considered significant. 

Because the SMA observations are not sensitive enough to detect the 
stellar photosphere, the alignment of the images from 
the SMA and Hubble Space Telescope is limited by the absolute astrometry. 
Even taking account of this uncertainty, it seems that the two millimeter
emission peaks do not align perfectly along the $30{\degr}$ position angle 
of the primary optical disk \citep{kal95}.
Instead, examination of Figure~\ref{fig:image} 
suggests that the peaks align more closely with the $34{\degr}$ position angle 
of the scattered light secondary disk described by \citet{gol06}. 
Observations with better resolution 
and sensitivity are needed to confirm this suggestion; 
the non-circular beam makes it difficult to assess small differences in 
orientation, and the millimeter emission structure itself may prove to be 
warped or more complex.

The 1.3~millimeter flux in the detected structure is $13\pm 1.4$~mJy, 
estimated by integrating over the emission in the image. This value is 
only approximately half of the $24.3\pm3.0$~mJy measured in the $24''$ 
SEST beam at 1.2~millimeters \citep{lis03}, a discrepancy significantly 
larger than expected from the mutual absolute calibration uncertainties 
and the spectral slope. The difference suggests the 
presence of an additional, extended 1.3~millimeter emission component, missed 
in these observations by the spatial filtering properties of the 
interferometer.  
Given the shortest SMA baselines, the peak brightness is diminished already 
by 50\% for a $20''$ (FWHM) Gaussian source \citep{wil94}, a size scale 
smaller than the SEST beam.
Judging from the partially resolved images from far-infrared and 
submillimeter filled aperture telescopes, 
this missing component is likely elongated along the disk, 
which extends beyond the SMA field of view.

The separate dust peak to the southwest is detected at 
$\sim5\sigma$, offset by ($-21\farcs4\pm0\farcs4,-22\farcs4\pm0\farcs6$) from 
the center of the field (not shown). The corresponding absolute position 
is $\alpha = 5^h47^m14\fs82$, $\delta =-51\degr04\arcmin21\farcs9$ (J2000).
Because the primary beam correction is large and uncertain at this location
beyond the half power point, it is difficult to provide an 
accurate estimate of the flux. The position is well determined, however, 
and shows that this peak does not lie along an extension of the optical 
disk. This supports previous suggestions that this feature is a background 
source, presumably a dusty galaxy.

\subsection{Belt Location and Width}
\label{sec:models}
\begin{figure}[h!]
\begin{center}
\includegraphics[scale=0.85,angle=0]{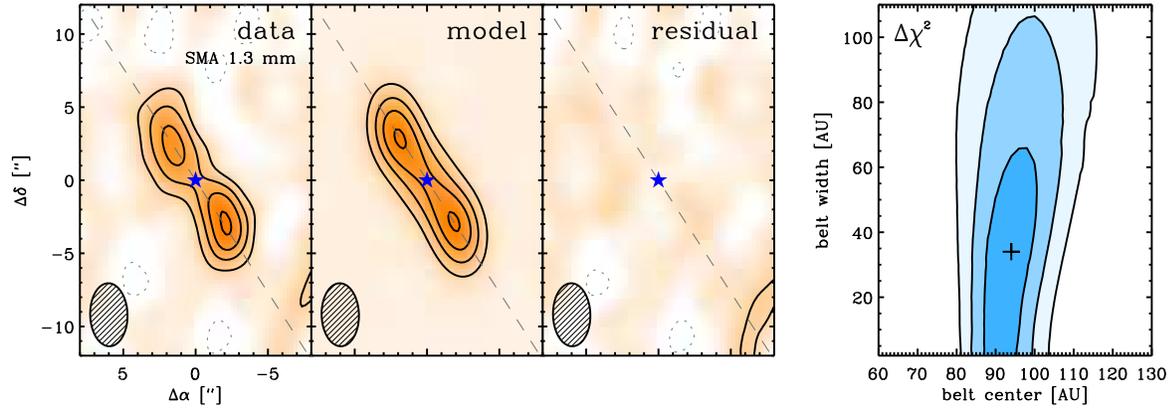}
\figcaption{
{\em Left panels:} SMA image of the 1.3 millimeter emission 
from $\beta$~Pictoris together with the image of the best-fit 
axisymmetric belt model and the residuals. The contour levels and
beam size are the same as in Figure~\ref{fig:image}. 
The dashed line indicates a position angle of $34{\degr}$.
{\em Right panel:} The $\chi^2$ surface for the belt center and width
model parameters, with contours at $1,2,3\sigma$. The cross marks the 
best-fit model.
\label{fig:model}}
\end{center}
\end{figure}

We constrain the basic properties of the 1.3 millimeter emission with 
a simple model that assumes the structure is characterized by a flat,
axisymmetric belt of emission. We take the radial profile of the emission
to be $r^{-0.5}$, which is physically motivated by optically thin emission 
for constant surface density 
and a temperature profile of $r^{-0.5}$ due to stellar irradiation.
We fix the inclination and orientation of the belt on the sky to
$87\degr$ and $33\degr$, respectively; due to the limited
resolution, small variations in these geometric parameters do not 
have significant effects on the results. This simple model has three 
parameters: 
the belt center $R$, 
belt width $\Delta R$, and flux $F$. We calculate a grid of models
over the parameter ranges
$60 < R < 130$~AU and $2 < \Delta R < 110$~AU in steps of $2$~AU, 
and $10 < F < 17$~mJy in steps of 0.5~mJy, and calculate $\chi^2$ values 
for each model using all of the SMA visibilities. 
The right panel of Figure~\ref{fig:model} shows the resulting 
$\chi^2$ surface (marginalized over the parameter $F$). 
The cross marks the best fit at
$R=94\pm 8$~AU, $\Delta R= 34^{+44}_{-32}$~AU, $F=15\pm2$~mJy;
the uncertainties represent the formal $1 \sigma$ errors.
The data strongly constrain the belt center location and allow for widths 
up to sizes comparable to the resolution of the observations.
The three left panels of Figure~\ref{fig:model} show the 1.3 millimeter 
image from 
Figure~\ref{fig:image} together with images of the best-fit model 
and the residuals, all made in the same way. 
The model reproduces the main features of the data, and the residuals 
are consistent with noise.
If a steeper radial emissivity were assumed, then the outer edge of the 
emission could extend further. However, a belt with a width that encroaches
much closer to the star than the best fit may be difficult to reconcile 
with the mid-infrared emission from the system. A proper model that considers 
the constraints of the full spectral energy distribution requires many 
more assumptions than made here, in particular about the grain composition, 
grain size distribution, collisional behaviors and dynamics.

\section{Discussion}
\label{sec:discussion}
The new millimeter observations improve substantially on previous 
single dish images and start to resolve fine structure in the 
$\beta$~Pictoris disk. Since large grains cannot travel far from 
their place of origin due to short grain-grain collisional timescales 
and negligible radiation effects, the emission at this long wavelength 
should trace the dust producing planetesimals.  Inspection and analysis of 
the resolved millimeter emission suggest a highly inclined ring or belt 
centered at a radius within or near $\sim95$~AU. Aside from the nearly 
edge-on viewing geometry, 
the millimeter morphology is strikingly similar to other well-studied 
A-type stars with substantial circumstellar dust, in particular 
Vega \citep[e.g.][]{mar06} 
and Fomalhaut \citep[e.g.][]{hol03}. The region interior to this 
belt in $\beta$~Pictoris is clearly not empty, as evidenced by 
mid-infrared imaging and spectroscopy \citep[e.g.]{tel05,che07}, but it 
must be relatively deficient in dust mass or millimeter-sized grains or both.

The location of the millimeter emission belt corresponds closely to a 
prominent break in the slope of the optical scattered light, as well as 
a change in the optical color gradient \citep{gol06}. These properties 
are plausibly explained in a model with dust producing planetesimals 
located just interior the break, with stellar radiation (and possibly 
also a stellar wind) creating a radial gradient in grain size 
\citep{aug01,str06}. 
In this scenario, the extended halo of emission along the disk plane 
would be dominated by a population of small grains blown out onto highly 
elliptical or hyperbolic orbits, possibly with temperatures above the 
local blackbody values, that cover a large area on the sky and give rise
to the fraction of millimeter emission missed by the interferometer.

One implication of the multi-component emission structure is that the 
far-infrared to millimeter spectral index of $2.34\pm0.07$ indicated by 
the integrated spectrum \citep{van10} may not be representative of any 
of the individual components. Depending on the details, it is possible,
for example, that the belt component could show a steeper spectral index 
that would be closer to expectations for a steady-state collisional 
cascade, without resorting to unusual fragmentation prescriptions or 
wavy grain size distributions \citep{the07}.  Until multi-wavelength 
observations are available in this regime that clearly resolve the relevant 
structures, it will be problematic to use the integrated spectral index to 
make conclusive inferences about the grain properties and size distributions.

The 1.3~millimeter image of the dust belt around $\beta$~Pictoris sets 
the stage for much improved future observations with the Atacama Large 
Millimeter/Submillimeter Array (ALMA), now under construction in Chile
(and much better placed for aperture synthesis observations of this 
southern source). More detailed millimeter images have the potential to 
determine, e.g. if the dust belt center is offset from the star, or if the 
emission exhibits pericenter glow or other asymmetries that could point to 
dynamical perturbations from additional planets in this remarkable system.

\acknowledgments{
We thank the SMA staff for scheduling and executing the 
two filler tracks that provided the data used in this paper. 
We also thank Sally Heap and Don Linder for providing the 
Hubble Space Telescope/STIS coronographic image in FITS format.
A.M.H. is supported by a fellowship from
the Miller Institute for Basic Research in Science.
}

{\it Facility:} \facility{Submillimeter Array}


\end{document}